\documentclass[prb,twocolumn,showpacs,superscriptaddress,amsmath,amssymb,reprint]{revtex4-1}

\usepackage{graphicx}
\usepackage{latexsym}
\usepackage{amsmath}
\usepackage{amssymb}
\usepackage{amsfonts}
\usepackage{color}
\usepackage{bm}

\begin{document}
\bibliographystyle{apsrev}

\title{Spin-valve effect in S/F$_1$/F$_2$ and F$_1$/S/F$_2$ structures of atomic thickness}

\author{Zh. Devizorova}
\affiliation{Moscow Institute of Physics and Technology, 141700 Dolgoprudny, Russia}
\affiliation{Kotelnikov Institute of Radio-engineering and Electronics RAS, 125009 Moscow, Russia}
\author{S. Mironov}
\affiliation{Institute for Physics of Microstructures, Russian Academy of Sciences, 603950 Nizhny Novgorod, GSP-105, Russia}

\date{\today}
\begin{abstract}
We show that the peculiarities of the electron band structure strongly affect the spin-valve effect in heterostructures consisting of a superconductor (S) and two ferromagnetic layers F$_1$ and F$_2$. For the S/F$_1$/F$_2$ systems the energy shift between the electron bands in the S and F$_2$ layers determines wether the critical temperature $T_c$ of the superconductor increases or decreases as a function of the angle $\theta$ between the magnetic moments in the ferromagnets. In the case of half-metallic F$_2$ layer the type of the spin-valve effect becomes dependent on the position of the only occupied spin-band and the minimum of $T_c$ may correspond to the anti-parallel, parallel or non-collinear orientations of magnetic moments. Also, for the first time, we analytically demonstrate the possibility of the triplet spin-valve effect in the F/S/half-metal structures.
\end{abstract}


\maketitle

\section{Introduction}
In spintronic devices the electron spin carries information instead of charge. This offers the opportunity to increase the data processing speed and decrease the power consumption in comparison with conventional semiconductor devices.\cite{Wolf_Science} The basic idea beyond spintronics is to align a spin and measure some quantity which depends on the degree of alignment in a predictable way. This requires the conservation of the spin polarization for the time scales exceeding the inverse operating frequency. However, in the diffusive semiconducting structures the typical spin life-time is short due to the spin-orbit coupling which limits their applicability.\cite{Linder_NatPhys}

During the past decade the focus of the research activity is shifting towards the concept of superconducting spintronics. It combines the advantages of the spin-based electronics and the extremely large life-time of quasiparticles in superconductors to design the novel ultra-fast and energy effecient operating elements for the computational devices.\cite{Linder_NatPhys,Eschrig_Rev} Moreover, the sensitivity of the superconductors to the external magnetic field or the exchange field in the adjacent ferromagnetic layer provides various effective mechanisms for the magnetic control of such superconducting spintronics elements. In particular, colossal magnetoresistance of artificial multilayered structures containing an s-wave superconductor (S) and several ferromagnetic (F) layers, allows to realize the spin-valve, the magnetically driven superconducting transistor.\cite{Oh_APL, Tagirov_PRL, Buzdin_EPL99} The critical temperature $T_c$ of S/F/F and F/S/F spin valves is known to depend strongly on the angle $\theta$ between the magnetic moments in the ferromagnets. Setting the temperature between the minimum and maximum of $T_c$ and changing the mutual orientation of the magnetic moments one can switch the system from the normal to the superconducting state. Thus, the resistivity of the structure can be significantly varied by changing the magnetizations in the ferromagnets.

However, the behavior of $T_c$ as a function of $\theta$ appears to be very sensitive to the system parameters. Even the small variations of the layers thickness, the diffussion coefficients or the exchange fields values produce qualitative changes in the dependence $T_c(\theta)$.\cite{Fominov_JETPL, Wu_PRB, Mironov_PRB14} Thus, the design of the spintronics devices requires the detailed information about the influence of different factors on $T_c$.

Up to now the significant progress has been achieved in understanding of the physics behind the spin-valve effect.\cite{Oh_APL, Tagirov_PRL, Buzdin_EPL99, Fominov_JETPL, Buzdin_EPL03, Tollis_PRB, Wu_PRB, Baladie_PRB, You_PRB, Bozovic_EPL, Halterman_PRB, Linder_PRB, Fominov_JETPL03, Lofwander_PRB, Mironov_PRB14} There are three main factors affecting the behavior of $T_c$ as a function of $\theta$: (i) the average exchange field in the structure, (ii) the interference effects, and (iii) the presence of spin-triplet correlations. First, the large average exchange field in two ferromagnets should damp the system critical temperature making the antiparallel (AP) orientation of magnetic moments more favorable for the superconductivity nucleation compared to the parallel (P) one.\cite{Tagirov_PRL, Buzdin_EPL99, Baladie_PRB, You_PRB, Buzdin_EPL03, Tollis_PRB, Bozovic_EPL, Halterman_PRB, Linder_PRB, Lofwander_PRB} At the same time, the exchange field splits the Fermi surfaces for spin-up and -down electrons which produces the spatial oscillations of the Cooper pair wave function inside the ferromagnets accompanied by the generation of spin-0 triplet correlations and gives rise to the interference phenomena which affect $T_c$. In particular, depending on the ratio between the thicknesses of the F-layers and the oscillation period the switching between P and AP orientations results in the enhancement or damping of $T_c$.\cite{Buzdin_RMP, Fominov_JETPL, Mironov_PRB14} The combination of the two described effects determines wether the maximum of $T_c$ corresponds to $\theta=\pi$ (standard spin valve effect) or $\theta =0$ (inverse spin valve effect).\cite{Fominov_JETPL, Mironov_PRB14} The situation becomes even more interesting if one considers the non-collinear orientation of magnetic moments. The non-collinearity results in the appearance of the spin-1 triplet components of the Green function inside all three layers.\cite{Bergeret_PRL} These components provide an additional channel for the leakage of Cooper pairs from the superconductor and, thus, decrease the critical temperature. As a result, the dependence $T_c(\theta)$ can become non-monotonic with a minimum at $\theta \ne 0,\pi$ (triplet spin valve effect).\cite{Fominov_JETPL, Wu_PRB, Mironov_PRB14}

The type of the spin valve effect appears to depend strongly on the relative  position of S and F layers. For S/F/F structures the theoretical analysis shows that in the dirty limit all the standard, inverse and triplet spin valve effects are possible.\cite{Fominov_JETPL} In the clean limit the numerical simulations based on the Bogoliubov--de Gennes equations also demonstrate the possibility of the triplet switching, however for all considered parameters only the case $T_c^{AP}>T_c^{P}$ was realized.\cite{Wu_PRB} Note that the theoretical results are well reproduced in the experiments where for different types of S/F/F hybrids the standard \cite{Nowak_SST, Nowak_PRB}, inverse \cite{Leksin_PRL11, Leksin_PRL12, Leksin_PRB} and triplet \cite{Leksin_PRL12, Zdravkov_PRB13} switching was observed.

For the spin valves of the F/S/F type the situation is different. For such structures the possible types of the spin valve effect depend not only on the concentration of impurities but also on the symmetry between the parameters of two ferromagnets (thicknesses, diffusion coefficients or exchange field values).  The majority of theoretical papers predict that for symmetric F/S/F structures only the case $T_c^P<T_c^{AP}$ is possible both in dirty \cite{Tagirov_PRL, Buzdin_EPL99, Baladie_PRB, You_PRB} and clean \cite{Bozovic_EPL, Halterman_PRB, Linder_PRB} limits. Moreover, in the dirty limit $T_c$ was found to be monotonically increasing for $0<\theta<\pi$ irrelevant to the symmetry or asymmetry of the F layers.\cite{Fominov_JETPL03, Lofwander_PRB} In contract, for the clean asymmetric F/S/F structures both the standard and inverse spin valve effect is possible, while in the symmetric case only the standard switching is realized.\cite{Mironov_PRB14} These predictions are consistent with the results of the recent experiments.\cite{Nowak_PRB, Kinsey, Gu_PRL, Potenza_PRB, Westerholt_PRL, Moraru_PRL, Moraru_PRB, Kim, Luo_EPL, Zhu_PRL, Rusanov_PRB, Aarts, Steiner_PRB, Singh_APL, Singh_PRB, Leksin_JETPL} Interestingly, the experiments on the Au/Co/Nb/Co/IrMn/Co/Ta/Si show the possibility to have even the triplet switching for the F/S/F system.\cite{Floksta_PRB} Theoretically such behavior was found only for the asymmetric clean structures with specific choice of parameters and the amplitude of the corresponding dip on the dependence $T_c(\theta)$ was pretty small.\cite{Mironov_PRB14}

It is important to note that in the theoretical papers cited above it was assumed that the density of states (DOS) is the same in all three layers. Such assumption allows to use the quasiclassical approximation which significantly simplify the calculation of $T_c$ and obtain the transparent analytical results. The applicability of the quasiclassical approach requires the negligibility of the exchange field and the energy shifts of the electron energy bands in different layers compared to the Fermi energy. Typically, the weak ferromagnets such as CuNi or PdFe which are often used in the experiments on the superconducting proximity effect satisfy these requirements.\cite{Oboznov_PRL, Bolginov_JETPL, Zdravkov_PRB13} However, the quasiclassical approach obviously fails when one deals with the strong ferromagnets (such as Co or CrO$_2$) \cite{Keizer_Nat, Anwar_PRB, Singh_PRX} or the combination of materials with very different electron band structure. Previously, both situations were shown to result in a dramatic changes of the critical temperature behavior. In particular, the changes in the electron band structure can change the standard spin-valve effect to the inverse one \cite{Montiel_EPL} while the strong spin polarization inside the strong ferromagnets enhances the triplet spin-valve effect \cite{Mironov_PRB15, EschrigLinder, Halterman2016} in contrast with the case of weak ferromagnets. Thus, to understand the full picture of the spin-valve effect in S/F/F and F/S/F systems it is crucial to go beyond the quasiclassical approximation and account the band features and the renormalization of the DOS due to the strong exchange field.

The attempts to develop the adequate quasiclassical theory describing the superconducting proximity effect with strong ferromagnets (often called half-metals) were performed within several approaches.\cite{Eschrig_NJP, Moor_PRB, Mironov_PRB15,EschrigLinder} The distinctive feature of half-metals (HM) is the strong energy splitting of the spin-up and spin-down electron bands which can exceed the Fermi energy.\cite{Pickett_PT, Coey_JAP} In such materials the DOS for electrons with the opposite spin projections $s_z$ to the quantization axis is substantially different. The simplest situation is realized if one assumes that only one spin band in HM is occupied while the DOS for another one is zero. This allows to introduce the quasiclassical Green function inside the HM and describe the proximity effect, e.g., using the Usadel theory. The full spin polarization results in the survival of only the triplet correlations while all other components of the Green function become fully suppressed.\cite{Mironov_PRB15} However, the consideration of more general case of two occupied spin bands within the quasiclassical approximation meets several fundamental problems. In particular, the process of matching of the Green functions inside and outside the HM becomes non-trivial (see [\onlinecite{Eschrig_NJP}] and references therein). Recently, the generic boundary conditions for the Usadel equation were derived and successfully applied for the description of the Josephson current through HM layer\cite{Eschrig_NJP} and the spin-valve effect.\cite{EschrigLinder} However up to now there are no clear results on the behavior of $T_c$ for the spin-valve structures with the HM for the arbitrary relative position of the spin energy bands. Alternatively, the Bogoluibov-de Geenes (BdG) \cite{Asano_PRL, Asano_PRB, Sawa, Beri_PRB,Halterman2016} or Blonder-Tinkham-Klapwijk \cite{Zheng_JPCM, Feng_PRB, Linder_PRB10, Enoksen_PRB} approaches were used to avoid the problem with the boundary conditions but the complexity of calculation does not allow to obtain the transparent analytical results while the applicability of the numerical ones is limited by the ranges of the system parameters under consideration. Thus, up to now there is no theory describing the spin-valve effect in the system with arbitrary strength of the exchange field inside the ferromagnets and for the arbitrary relative position of the electron energy spin-bands.

At the same time one can expect that the strong spin polarization of the superconducting correlations inside HM can strongly enhance the magnitude of the triplet spin-valve effect in the S/F/HM systems.\cite{Singh_PRX, Mironov_PRB15, Moor_PRB, EschrigLinder, Halterman2016} In contrast to the usual S/F/F structures where the triplet switching is typically washed out by the singlet and spin-0 triplet correlations, in dirty S/F/HM systems the dependence $T_c(\theta)$ is determined only by the spin-1 triplet correlations. This leads to the symmetry $T_c(\pi-\theta) = T_c(\theta)$ and the substantial increase in the variations of $T_c$ when varying the angle $\theta$. At the same time, the analysis of the influence of the strong spin polarization in the HM on the spin-valve effect in the situation when the superconductor is placed between the ferromagnet and HM is still lacking. Also it is not clear how the position of the electron spin bands inside the HM layer affects the type of the spin-valve effect. Similar to the F/S/F systems \cite{Montiel_EPL} one can expect that the shift or inversion of the spin bands can result in the qualitative changes in the behavior of $T_c$.

In the present paper we propose the theory of the spin valve effect in the S/F$_1$/F$_2$ and F$_1$/S/F$_2$ structures of atomic thickness. We use the combination of the Gor'kov formalism and the tight-binding model, which was previously applied for the description of the proximity effect in RuSr$_2$GdCu$_2$O$_8$ and La$_{0.7}$Ca$_{0.3}$MnO$_3$ compounds.\cite{Andreev_PRB, Buzdin_EPL03, Tollis_PRB, Prokic_PRB} Our model enables the exact analytical solution which is applicable for arbitrary magnitudes of the exchange field (both for weak ferromagnets and half-metals) and arbitrary relative position of the electron energy bands in different layers. We demonstrate that for the S/F$_1$/F$_2$ systems the shift between the electron spin-bands in the F$_2$ layer relative to the ones in the superconductor results in the crossover from the standard to inverse spin-valve effect. Moreover, in the case when the F$_2$ layer is half-metallic the form of the dependence $T_c(\theta)$ strongly depends on the position of the occupied spin-band. Specifically, if the position of this band coincides with the one in the S layer the spin-valve effect is purely triplet and $T_c(0)=T_c(\pi)$ in a qualitative agreement with the results obtained within the Usadel approach. However, the shift of the energy band in the HM layer results in the asymmetry of $T_c(0) \ne T_c(\pi)$ and depending on the shift the standard or inverse switching becomes possible.

Also for the first time we analytically demonstrate the possibility of the triplet switching for the spin valves of the F$_1$/S/F$_2$ type. Previously such behavior of $T_c$ was described only using numerical calculations \cite{Mironov_PRB14}. It was pointed out that the key role in the emerging of the triplet switching is played by the asymmetry of the structure. Our calculations confirm this conclusion and show that such unusual behavior of $T_c$ is possible provided one of the ferromagnetic layers is half-metallic while the other one is not.

The paper is organized as follows. In Sec. \ref{Sec:SF1F2} we consider the S/F$_1$/F$_2$ spin valve with arbitrary exchange field values in the F$_2$ layer and analyze the effect of the band structure on the superconducting critical temperature. In Sec. \ref{Sec:SFHM} the detailed analysis of the spin-valve effect in F/S/HM structure is provided. In Sec. \ref{Sec:Conclusion} we summarize our results.

\section{Spin valve effect in S/F1/F2 structure} \label{Sec:SF1F2}

\begin{figure}[t!]
\includegraphics[width=0.3\textwidth]{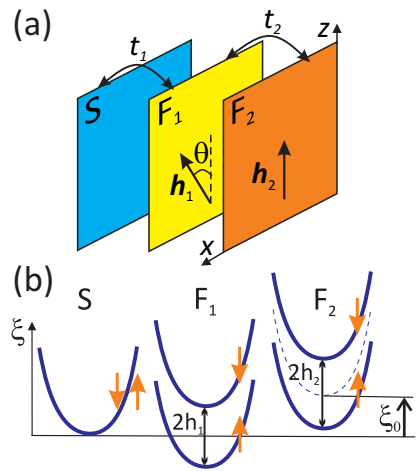}
\caption{(Color online) The S/F$_1$/F$_2$ spin valve of atomic thickness. (a) The sketch of the structure. Here $\theta$ is the angle between the exchange fields in the ferromagnets. The adjacent layers are coupled by the transfer integrals $t_1$ and $t_2$, respectively. (b) The electron energy band structure. The parameter $\xi_0$ is the energy shift between the band of the F$_2$ ferromagnet in the absence of the exchange field and the electron energy band in the superconductor.} \label{Fig2}
\end{figure}

In this section we calculate the critical temperature of the S/F$_1$/F$_2$ trilayer of atomic thickness, which is schematically shown in Fig.\ref{Fig2}(a). The $y$-axis is chosen to be perpendicular to the layers' interfaces. The exchange field ${\bf h}_2$ in the F$_2$ layer is parallel to the $z$-axis, while the exchange field ${\bf h}_1$ in the F$_1$ layer belongs to the $xz$-plane and forms the angle $\theta$ with the $z$-axis: ${\bf h}_1=h_1\cos \theta \hat{\bf z}+h_1\sin \theta \hat{\bf x}$.

We assume the quasiclassical electron motion characterized by the momentum inside the layers, while the quasiparticles motion perpendicular to the layers is described by the tight-binding model. The transfer integrals $t_1$ between S and F$_1$ layers and $t_2$ between F$_1$ and F$_2$ ones are assumed to be much smaller than $T_c$. Also we restrict ourselves to the limit of coherent interlayer tunneling which conserves the in-plane electron momentum.

To account only the most important features of the electron band structure in the system we assume that in the absence of the exchange field in the ferromagnets the in-plane quasiparticles motion is described by the energy spectrum $\xi({\bf p})$ in the S and F$_1$ layers, while in the F$_2$ layer the electron energy band is shifted by $\xi_0$ with respect to the energy band of the superconductor, i.e the spectrum is $\xi({\bf p})+\xi_0$ with the arbitrary sign of $\xi_0$ [see Fig.\ref{Fig2}(b)]. The exchange field splits the spin-up and spin-down bands in the F$_1$ and F$_2$ layers by the distance $2h_1$ and $2h_2$, respectively. By varying $h_2$ and $\xi_0$ the system can be switched from S/F/F to S/F/HM regimes.

\subsection{Analytical model}

To calculate the critical temperature of the superconducting layer we use the Gor'kov formalism. Let the electron annihilation operators in the S, F$_1$ and F$_2$ layers are $\phi$, $\psi$ and $\eta$, respectively. Then the Hamiltonian of the system reads
\begin{equation}
\label{H}
\hat H=\hat H_0+\hat H_S+\hat H_t.
\end{equation}
Here the first term
\begin{equation}
\label{H0}
\hat H_0=\sum \limits_{{\bf p};\alpha,\beta=\{1,2\}} \left(\xi({\bf p})\phi^+_{\alpha}\phi_{\beta} \delta_{\alpha \beta}+\hat A_{\alpha \beta}\psi^+_{\alpha}\psi_{\beta}+\hat B_{\alpha \beta}\eta^+_{\alpha}\eta_{\beta}\right)
\end{equation}
describes the electron motion in each isolated layer in the normal state, the term
\begin{equation}
\label{Hs}
\hat H_S=\sum \limits_{{\bf p}}\left(\Delta^*  \phi_{{\bf p},2}\phi_{-{\bf p},1}+\Delta  \phi^+_{{\bf p},1}\phi^+_{-{\bf p},2}\right)
\end{equation}
stands for the s-wave Cooper pairing in the superconductor, and the last term
\begin{equation}
\label{Ht}
\hat H_t=\sum \limits_{{\bf p};\alpha=\{1,2\}}\left[t_1( \phi^+_{\alpha}\psi_{\alpha}+\psi^+_{\alpha}\phi_{\alpha})+t_2(\psi^+_{\alpha}\eta_{\alpha}+\eta^+_{\alpha}\psi_{\alpha})\right]
\end{equation}
describes the tunneling between the layers. In Eqs.~(\ref{H0})-(\ref{Ht}) $\alpha$ and $\beta$ are the spin indexes. The form of the matrices $\hat A$ and $\hat B$ accounts the Zeeman interaction and different possible orientations of the magnetic moments in the F$_1$ and F$_2$ layers, respectively:
\begin{equation}
\label{A}
\hat A=\left( \begin{array}{cc}
		\xi({\bf p})-h_1\cos \theta & -h_1\sin \theta\\
		-h_1\sin \theta & \xi({\bf p})+h_1\cos \theta
	\end{array} \right),
\end{equation}
\begin{equation}
\label{B}
\hat B=\left( \begin{array}{cc}
		\xi({\bf p})+\xi_{0}-h_2 & 0\\
		0 & \xi({\bf p})+\xi_{0}+h_2
	\end{array} \right).
\end{equation}

In the imaginary-time representation the minimal set of the Green functions required for the critical temperature calculation is
\begin{equation}
G_{\alpha,\beta}=-\langle T_{\tau}(\phi_{\alpha},\phi_{\beta}^+)\rangle, \qquad F^{+}_{\alpha,\beta}=\langle T_{\tau}(\phi_{\alpha}^+,\phi_{\beta}^+)\rangle,
\end{equation}
\begin{equation}
E^{\psi}_{\alpha,\beta}=-\langle T_{\tau}(\psi_{\alpha},\phi_{\beta}^+)\rangle, \qquad F^{\psi+}_{\alpha,\beta}=\langle T_{\tau}(\psi_{\alpha}^+,\phi_{\beta}^+)\rangle,
\end{equation}
\begin{equation}
E^{\eta}_{\alpha,\beta}=-\langle T_{\tau}(\eta_{\alpha},\phi_{\beta}^+)\rangle, \qquad F^{\eta+}_{\alpha,\beta}=\langle T_{\tau}(\eta_{\alpha}^+,\phi_{\beta}^+)\rangle.
\end{equation}
Taking the imaginary-time derivatives of all Green functions in the Fourier representation and using the Heisenberg equations for the operators $\phi$, $\psi$ and $\eta$ with the Hamiltonian (\ref{H}), we obtain the system of Gor'kov equations:
\begin{gather}
(i\omega-\xi)G+\Delta I F^{+}-t_1E^{\psi}=\hat 1, \\
(i\omega+\xi)F^+ -\Delta^* I G+t_1F^{\psi+}=0, \\
(i\omega-\hat A)E^{\psi}-t_1 G -t_2 E^{\eta}=0, \\
(i\omega+\hat A)F^{\psi+}+t_1 F^+ + t_2 F^{\eta+}=0, \\
(i\omega-\hat B)E^{\eta}-t_2 E^{\psi}=0, \\
(i\omega+\hat B)F^{\eta+}+t_2 F^{\psi+}=0.
\end{gather}

The critical temperature $T_c(\theta)$ of the system is determined by the self-consistency equation
\begin{equation}
\label{SCE}
T_c(\theta)=T_c(0)-T_{c0}^2 \sum \limits_{\omega_n=-\infty}^{+\infty} \int \limits_{\xi=-\infty}^{+\infty} d\xi \frac{\hat F^+_{12}(\theta)-\hat F^+_{12}(0)}{\Delta^*},
\end{equation}
where $T_{c0}$ is the critical temperature in the absence of the proximity effect, i.e. $t_1=t_2=0$, $\omega_n=\pi T_{c0}(2n+1)$ are the Matsubara frequencies, and the anomalous Green function $\hat F^+$ is calculated only in the first order of the perturbation theory with the gap potential as a small parameter.

The straightforward solution of the Gor'kov equations for the anomalous Green function ${\hat F^+}$ reads
\begin{multline}
\label{AGF1}
\frac{\hat F^+}{\Delta^*}=\biggl\{(i\omega +\xi)\hat 1 -t_1^2\biggl[(i\omega +\hat A)-t_2^2(i\omega +\hat B)^{-1}\biggr]^{-1}\biggr\}^{-1} \\ \times \hat I \biggl\{(i\omega -\xi)\hat 1-t_1^2\biggl[(i\omega -\hat A)-t_2^2(i\omega -\hat B)^{-1}\biggr]^{-1} \biggr\}^{-1},
\end{multline}
where $\hat I=i \sigma_y$. The substitution of Eq.~(\ref{AGF1}) into Eq.~(\ref{SCE}) gives the desired critical temperature of the spin valve.

\subsection{Results and discussion}

In this section we show how the peculiarities of the band structure of the F$_2$ ferromagnet influence the critical temperature of the superconductor. In what follows for simplicity we assume the tunneling constants $t_1$ and $t_2$ to be small and expand Eq.~(\ref{AGF1}) over these parameters keeping the terms up to the sixth order (see Appendix \ref{App:AGF_SF1F2} for details).

After this simplification in all possible cases the system critical temperature can be represented in the form
\begin{equation}
\label{Tc_gf}
T_c(\theta)=T_c(0)+a(1-\cos \theta)+b \sin^2 \theta,
\end{equation}
where $a$ and $b$ are the coefficients which determine the behavior of $T_c$ as a function of $\theta$ and depend on the parameters of the system, $T_c(0)$ is the critical temperature at the angle $\theta=0$. The sign of $a$ defines whether the standard or inverse spin valve effect is realized in the system: if $a>0$ ($a<0$) the maximum in $T_c(\theta)$ corresponds to $\theta=\pi$ ($\theta=0$). The possibility of the triplet spin valve effect depends on an interplay between the values of $a$ and $b$, i.e. the minimum in $T_c(\theta)$ corresponds to the non-collinear orientation of the exchange fields if $|a|<2|b|$ and $b<0$. Below we calculate $a$ and $b$ for different limiting cases and analyze the effect of the electron band structure of the F$_2$ ferromagnet on these coefficients and, thus, on the type of the spin-valve effect.

Substituting the expansion of $\hat F^+$ into (\ref{SCE}) and integrating over $\xi$ we obtain the expression for $a$ and $b$:
\begin{equation}
\label{a_SFF_gf}
a=\sum \limits_{\omega_n>0}\frac{8\pi T_{c0}^2 t_1^2 t_2^2 h_1 h_2(4\omega^2+h_+h_-)}{\omega p^2 a_+b_+} \left[1+O(t^2) \right],
\end{equation}
\begin{equation}
\label{b_SFF_gf}
b=\sum \limits_{\omega_n>0}\frac{16\pi T_{c0}^2 t_1^2 t_2^4 h_1^2 h_2^2 (a_- b_-+16\omega^2h_+h_-)}{\omega p^3 a_+^2 b_+^2},
\end{equation}
where $h_{\pm}=(h_2 \pm \xi_0)$, $a_{\pm}=(h_-^2 \pm 4\omega^2)$, $b_{\pm}=(h_+^2 \pm 4\omega^2)$, $p=(4\omega^2+h_1^2)$.

The expressions (\ref{a_SFF_gf}) and (\ref{b_SFF_gf}) are applicable both for weak and strong F$_2$ ferromagnet. Let us analyze these two cases separately.

\subsubsection{F$_2$ layer: the limit of weak ferromagnet}\label{Sec_weak}

We begin with the limit when in the S/F$_1$/F$_2$ spin valve both ferromagnets are weak. For simplicity we put $h_1=h_2=h$ and $t_1=t_2=t$. Aiming to compare our results with quasiclassical ones qualitatively, we first consider the S/F/F spin valve with two identical ferromagnets, i.e. $\xi_0=0$. In this case
\begin{equation}
a=\sum \limits_{\omega_n>0}\frac{8\pi T_{c0}^2 t^4 h^2 }{\omega (4\omega^2+h^2)^3}>0, \qquad b\ll a.
\end{equation}
One sees that the critical temperature is a monotonically increasing function of the angle $\theta$ and, thus, the trilayer demonstrates only the standard spin valve effect. In the quasiclassical approach the interference effects due to the finite layers thickness and the typical scale of the oscillations of the anomalous Green function in the ferromagnets can lead to the inverse and triplet switching. In the model with atomically thin layers such interference effects are not possible, and only the influence of the average exchange field on $T_c$ is sufficient.

\begin{figure}[t!]
\includegraphics[width=0.3\textwidth]{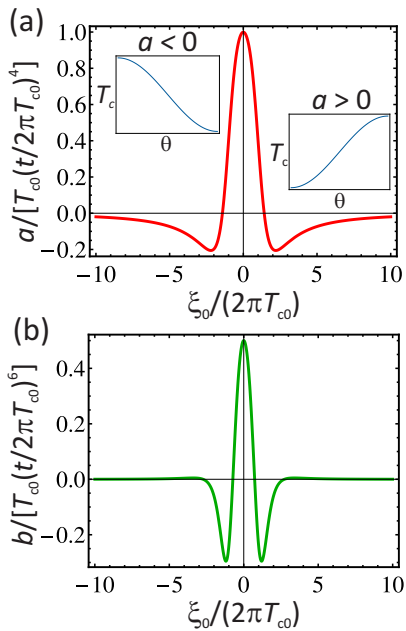}
\caption{(Color online) The critical temperature can be represented as $T_c(\theta)=T_c(0)+a(1-\cos \theta)+b \sin^2 \theta$. (a) The coefficient $a$ vs. the energy separation $\xi_0$ at $h_1=2\pi T_{c0}$. The insets show the corresponding dependencies of the critical temperature on the angle $\theta$. (b) The same for the coefficient $b$.} \label{Fig3}
\end{figure}

Next we analyze the effect of the electron band structure in the F$_2$ ferromagnet on the critical temperature, which is beyond the quasiclassical approximation. For this purpose we introduce the finite energy separation between the conduction band minimum without the exchange field in F$_2$ layer and the conduction band minimum in the superconductor, i.e. $\xi_0 \ne 0$. The dependence of $a$ and $b$ on $\xi_0$ are presented in Fig.\ref{Fig3}(a) and Fig.\ref{Fig3}(b), respectively. These dependencies have several important features. First, there is the range of $\xi_0$, where $a<0$ giving rise to the inverse spin valve effect. Thus, not only the interference effects, but also the band structure details, which cannot be taken into account in the quasiclassical equations, result in inverse switching in S/F$_1$/F$_2$ spin valve. Second, the coefficient $b$ can be positive. However, we emphasize that the positive $b$ coefficient do not lead to the appearance of the local maximum in $T_c$, because we used the Taylor expansion over $t$ and $b$ always smaller than $a$.

\subsubsection{F$_2$ layer: the limit of half-metal}

The results of Sec.~\ref{Sec_weak} show that in our model of the S/F$_1$/F$_2$ structures with two weak ferromagnets the triplet spin valve effect is suppressed. However, the situation changes if one replaces the weak ferromagnet F$_2$ with a half-metallic layer. In this case the full spin polarization inside the half-metal leads to the great enhancement of the triplet spin valve effect. To demonstrate it explicitly, we now consider the S/F/HM spin valve.

\begin{figure}[t!]
\includegraphics[width=0.4\textwidth]{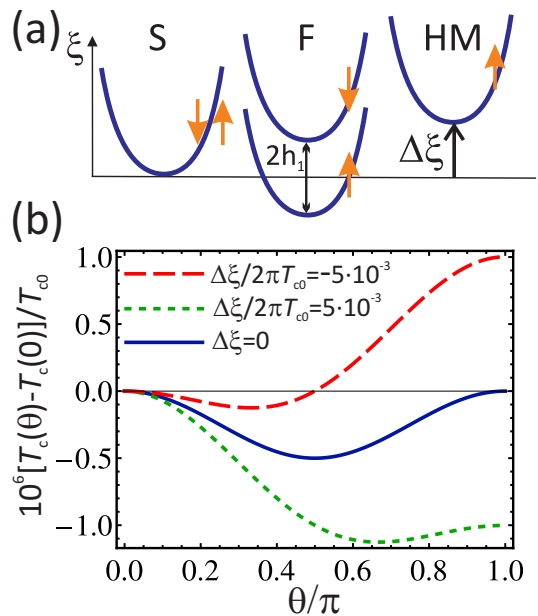}
\caption{(Color online) (a) The electron energy band structure of S/F/HM spin valve. The spin-up band in the half-metal is shifted by $\Delta \xi$ with respect to the band in the superconductor. (b) The dependence of the superconducting critical temperature $T_c$ on the angle $\theta$ for different energy separations $\Delta \xi$. The value $\Delta \xi=0$ corresponds to the pure triplet spin valve effect, while for small, but finite $\Delta \xi$ the minimum of $T_c$ is shifted from $\pi/2$ and $T_c(0) \ne T_c(\pi)$. The parameters are $h_1/(2\pi T_{c0})=1$, $t_1=t_2=t$, $t/(2\pi T_{c0})=0.1$} \label{Fig4}
\end{figure}

The band structure of the S/F/HM spin valve is shown in Fig.\ref{Fig4}(a). Due to the large Zeeman splitting in HM, we assume the energy spectrum for spin-down quasiparticles to be $\xi_{\downarrow}=+\infty$, while for spin-up ones $\xi_{\uparrow}=\xi({\bf p})+\Delta \xi$, where $\Delta \xi$ is the energy shift between the spin-up band in HM and the electron energy band in the S layer. To obtain $a$ and $b$ for this case, in (\ref{a_SFF_gf}) and  (\ref{b_SFF_gf}) we should set $h_2=+\infty$, $\xi_0=+\infty$, $\xi_0-h_2=\Delta \xi$.

First, we consider the case $\Delta \xi=0$, i.e. the spin-up band in the half-metal coincides with the superconductor band,
\begin{equation}
a=0, \qquad b=-\sum \limits_{\omega_n>0}\frac{\pi T_{c0}^2 t_1^2 t_2^4   h_1^2}{\omega^3 (4\omega^2+h_1^2)^3}.
\end{equation}
This result is qualitatively the same as the one obtained previously in Ref.[\onlinecite{Mironov_PRB15}]. The dependence of the critical temperature on the angle between magnetic moments is non-monotonic with a minimum corresponding to $\theta=\pi/2$ and $T_c(0)=T_c(\pi)$, which is pure triplet spin valve effect.

Interestingly, the triplet spin valve effect emerging in the quasiclassical approximation is suppressed with increasing the exchange field in the F$_2$ ferromagnet. However, if the F$_2$ layer is half-metallic, i.e.  ${\bf h}_2$ is strong, the effect is significantly enhanced. This contradiction is due to the quasiclassical theory of the proximity effect in S/F/F structures cannot be applied directly to S/F/HM spin valves. At the same time, the model with atomically thin layer is describe the superconducting proximity effect both with weak ferromagnets and half-metals. In spite of the effect of finite layers thickness resulting in the triplet spin valve effect in S/F/F structures can not be taken into account in our model, we obtain the enhancement of the triplet spin valve effect in S/F/HM due to the F$_2$ layer band structure. Thus, we have shown that there is an alternative mechanism leading to the triplet spin valve effect in S/F/F and S/F/HM structures.

Now we analyze the effect of the energy shift $\Delta \xi$ on $T_c$. We obtain:
\begin{equation}
a=-\sum \limits_{\omega_n>0}\frac{4\pi T_{c0}^2 t_1^2 t_2^2 h_1 \Delta \xi}{\omega (4\omega^2+h_1^2)^2(4\omega^2+\Delta \xi^2)},
\end{equation}
\begin{equation}
b=-\sum \limits_{\omega_n>0}\frac{4\pi T_{c0}^2 t_1^2 t_2^4   h_1^2(4\omega^2-\Delta \xi^2)}{\omega (4\omega^2+h_1^2)^3(4\omega^2+\Delta \xi^2)^2}.
\end{equation}
One sees, that the nonzero energy separation $\Delta \xi$ leads to to the suppression of the triplet spin valve effect. However, the dependence $T_c(\theta)$ can still be non-monotonic, if the shift is rather small, i.e. $|a|<2|b|$ (see Fig.\ref{Fig4}(b)). The coefficient $b$ riches the maximum $t_1^2 t_2^4 /T_{c0}^5$ at $h_1 \sim T_{c0}$. At the same time, $a$ at $h_1 \sim T_{c0}$ is proportional to $t_1^2 t_2^2 \Delta \xi /T_{c0}^4$. Thus, for example, the dependence is non-monotonic at $h_1 \sim T_{c0}$ if $|\Delta \xi| < 2t_2^2/T_{c0}$. However, the minimum is shifted from $\pi /2$ and the effect is not pure triplet, i.e. $T_c(0) \ne T_c(\pi)$.

\begin{figure}[t!]
\includegraphics[width=0.3\textwidth]{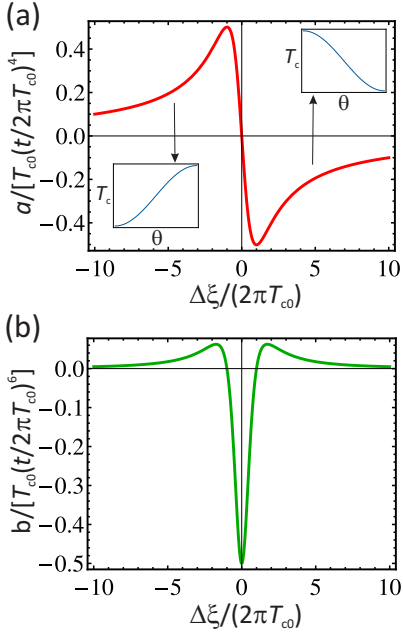}
\caption{(Color online) The dependencies of the $a$ and $b$ coefficients on the energy shift $\Delta \xi$ for the S/F/HM spin valve with $h_1/(2\pi T_{c0})=1$ are presented on the panels (a) and (b), respectively.} \label{Fig5}
\end{figure}

With the increase in the energy separation $|\Delta \xi|$ the dependence $T_c(\theta)$ becomes monotonic with $T_c(0) < T_c(\pi)$, if the spin-up band minimum is below the superconductor one, or $T_c(0) > T_c(\pi)$ in the other case. Indeed, depending on the sign of $\Delta \xi$ the coefficient $a$ can be both negative and positive giving rise to the standard and inverse spin valve effect, respectively [see Fig.\ref{Fig5}(a)].

The dependence of $b$ on the energy shift $\Delta \xi$ is presented in Fig.\ref{Fig5}(b). Despite the value $b$ can be positive, there is no maximum in $T_c$, because for such parameters the coefficient $a$ is nonzero, moreover $a \gg b$ due to the assumption made $t \ll T_{c0}$. However, the positive sign of $b$ may be a signature of possible maximum in $T_c$ if one consider more sophisticated model taking into account the not only the band shifts in different layers but also the finite layers thickness and the arbitrary transparencies of the interfaces.

Thus, we have shown that all the standard, inverse and triplet spin valve effects are possible in the S/F/HM structure with atomically thin layers depending on the position of the spin-up band in the half-metal with respect to the electron energy band in the superconductor.

\section{Spin valve effect in F/S/HM structure} \label{Sec:SFHM}

\begin{figure}[t!]
\includegraphics[width=0.3\textwidth]{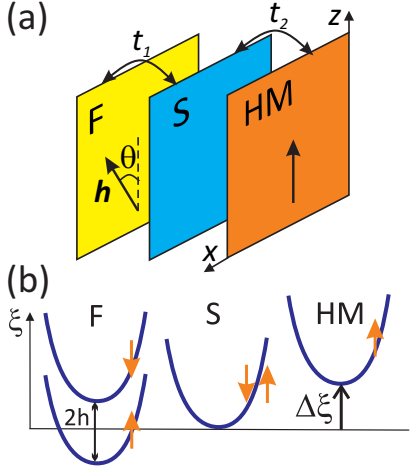}
\caption{(Color online) F/S/HM spin valve with atomically thin layers is schematically depicted on the panel (a). The exchange field in the ferromagnet forms the angle $\theta$ with spin quantization axes in the half-metal. The transfer integral $t_1$ couples ferromagnetic layer with the superconductor, while the superconductor and the half-metal are linked by $t_2$. (b) The band structure of the spin valve.} \label{Fig6}
\end{figure}

This section is devoted to the calculation of the critical temperature of the spin valve of the F/S/HM type where the ferromagnets are placed at different sides of the superconductor. This system is schematically shown in Fig.\ref{Fig6}(a). The spin quantization axis in the half-metal coincides with $z$-axis, while the exchange field ${\bf h}$ in the F layer has two components $h_z=h\cos\theta$, $h_x=h\sin\theta$. The transfer integral $t_1$ couples the S and F layers, while $t_2$ couples the S and HM. As before, in-plane quasiparticles motion in the S and F layers is described by the same energy spectrum $\xi({\bf p})$, while the energy spectrum in the half-metal is spin dependant: $\xi_{\uparrow}=\xi({\bf p})+\Delta \xi$ and $\xi_{\downarrow}=+\infty$ [see Fig.\ref{Fig6}(b)].

Below we calculate the critical temperature $T_c$ from the self-consistency equation (\ref{SCE}). To do this we find the anomalous Green function using the Gor'kov formalism. The electron annihilation operators in the S, F and HM layers are $\phi$, $\psi$ and $\eta$, respectively. In this notations the Hamiltonian still has the form (\ref{H}), with $\hat H_S$ and $\hat H_0$ satisfying (\ref{Hs}) and (\ref{H0}), respectively, but the part describing the tunneling between the layers becomes
\begin{equation}
\label{Ht2}
\hat H_t=\sum \limits_{{\bf p};\alpha=\{1,2\}}\left[t_1( \phi^+_{\alpha}\psi_{\alpha}+\psi^+_{\alpha}\phi_{\alpha})+t_2(\phi^+_{\alpha}\eta_{\alpha}+\eta^+_{\alpha}\phi_{\alpha})\right].
\end{equation}

The matrices $A$ and $B$ in $\hat H_0$ [see Eq.~(\ref{H0})] should be replaced by $\hat C$ and $\hat P$, respectively,
\begin{equation}
\hat C=\left( \begin{array}{cc}
		\xi({\bf p})-h\cos \theta & -h\sin \theta\\
		-h\sin \theta & \xi({\bf p})+h\cos \theta
	\end{array} \right),
\end{equation}
\begin{equation}
\hat P=\left( \begin{array}{cc}
		\xi({\bf p})+\Delta \xi & 0\\
		0 & +\infty
	\end{array} \right).
\end{equation}

We find the Fourier component of the anomalous Green function from the system of Gor'kov equations (see Appendix \ref{App:FSHM}):
\begin{multline}
\label{AGF2}
\frac{\hat F^+}{\Delta^*}=\left[(i\omega +\xi)\hat 1 -t_1^2(i\omega +\hat C)^{-1}-t_2^2(i\omega +\hat P)^{-1}\right]^{-1} \times \\ \times \hat I \left[(i\omega -\xi)\hat 1-t_1^2(i\omega -\hat C)^{-1}-t_2^2(i\omega -\hat P)^{-1}\right]^{-1}.
\end{multline}

To simplify the further calculations we again assume $t_1$ and $t_2$ to be small and perform the power expansion of (\ref{AGF1}) over these parameters. Then one should substitute the anomalous Green function (\ref{AGF2}) into the self-consistency equation (\ref{SCE}) to obtain $T_c(\theta)$.

First, we consider the limit $\Delta \xi=0$. In this case the non-trivial dependence of $T_c$ on the angle $\theta$ appears only in the eighth order. After some algebra we arrives to
\begin{multline}
T_c(\theta)=T_c(0) - \sum \limits_{\omega_n=-\infty}^{+\infty} \int \limits_{\xi=-\infty}^{+\infty} \frac{T_{c0}^2 h^2 t_1^4 t_2^4 \sin^2\theta d\xi}{\omega_+^4\omega_-(\omega_+^2-h^2)} \times \\ \times \biggl(-\frac{2}{\omega_+^3(\omega_+^2-h^2)} +\frac{1}{\omega_+\omega_-^2(\omega_+^2-h^2)}-\frac{1}{\omega_+^2\omega_-(\omega_-^2-h^2)}+\\+\frac{1}{\omega_-^3(\omega_-^2-h^2)}\biggr),
\end{multline}
where $\omega_+=i\omega+\xi$, $\omega_+=i\omega-\xi$. Integrating over $\xi$ we finally obtain:
\begin{equation}
T_c(\theta)=T_c(0)-\sum \limits_{\omega_n>0}\frac{\pi T_{c0}^2t_1^4 t_2^4  \sin^2\theta}{4\omega^7 h^2} P\left(\frac{\omega}{h} \right),
\end{equation}
where
$P(x)$ is the positively defined function (see Appendix \ref{App:FSHM}).

The dependence $T_c(\theta)$ clearly demonstrates the triplet spin valve effect, which is not suppressed by the correlations with $s_z=0$. This result is qualitatively similar to the ones in Ref.[\onlinecite{Mironov_PRB14}], where the possibility of the triplet spin valve effect in clean F$_1$/S/F$_2$ structures was found.

The magnitude of the effect can be characterized by the value $\delta T_c=T_c(0)-T_c(\pi/2)$, which riches the maximum $\delta T_c \propto t_1^4 t_2^4 / T_{c0}^7$ at $h \sim T_{c0}$, while both for $h \gg T_{c0}$ and $h \ll T_{c0}$ the triplet spin valve effect is small. Indeed, if $h \gg T_{c0}$ we find:
\begin{equation}
\frac{\delta T_c}{T_{c0}} = \gamma_1 \frac{t_1^4 t_2^4}{T_{c0}^8}\left(\frac{T_{c0}}{h}\right)^2 ,
\end{equation}
where $\gamma_1=635\zeta(7)/ (512\pi^6) \simeq 0.001$. In the opposite limit $h \ll T_{c0}$ we obtain
\begin{equation}
\frac{\delta T_c}{T_{c0}} = \gamma_2 \frac{t_1^4 t_2^4}{T_{c0}^8}\left(\frac{h}{T_{c0}}\right)^2,
\end{equation}
where $\gamma_2=26 611\zeta(11)/ (32768\pi^{10}) \simeq 9 \times 10^{-6}$.

Now we analyze the effect of the energy separation between bands, i.e. $\Delta \xi \ne 0$, on the $T_c$. In this case the dependence on the angle $\theta$ appears in the fourth order
\begin{multline}
T_c(\theta)=T_c(0)-\sum \limits_{\omega_n>0}\frac{2\pi T_{c0}^2t_1^2 t_2^2 h \Delta \xi (1-\cos \theta)}{\omega^7} K(x_1,x_2),
\end{multline}
where $x_1=h/\omega$, $x_2=\Delta \xi / \omega$ and the function $K(x_1,x_2)$ is explicitly written in Appendix \ref{App:FSHM}.

Similar to the case of S/F/HM structure the finite $\Delta \xi$ suppresses the triplet spin valve effect giving rise to the standard or inverse switching depending on the sign of $\Delta \xi$.

\section{Conclusion} \label{Sec:Conclusion}

In conclusion, we propose the theory of the spin valve effect in atomically thin S/F$_1$/F$_2$ and F$_1$/S/F$_2$ multilayered structures. Using the Gor'kov formalism we demonstrate that the dependence of the critical temperature $T_c$ on the angle $\theta$ between the exchange field vectors ${\bf h}_1$ and ${\bf h}_2$ is extremely sensitive to the peculiarities of the electron band structure of the F$_2$ ferromagnet. For the S/F$_1$/F$_2$ structure where both ferromagnets are weak and in the absence of the exchange field the electron energy spectra in the S and F$_2$ layers are shifted by some energy both the standard and inverse switching was shown to be possible depending on the value of this shift. The dependence $T_c(\theta)$ becomes even more sensitive to the band structure in the case when F$_2$ ferromagnet is strong, i.e. half-metallic. If the only occupied spin-band in the half-metal coincides with the band in the S layer, then the dependence $T_c(\theta)$ is non-monotonic with the minimum corresponding to $\theta=\pi /2$ and $T_c(0)=T_c(\pi)$, i.e. the spin valve effect is pure triplet in agreement with the results obtained with the Usadel theory \cite{Mironov_PRB15}. However, the changes in the position of the energy band in the half-metal result in the suppression of the triplet spin valve effect, shift of the minimum from $\theta=\pi /2$, and the asymmetry of $T_c(0) \ne T_c(\pi)$. Depending on the sign of the energy shift the maximum of $T_c(\theta)$ corresponds to $\pi$ or $0$ giving rise to the standard and inverse spin valve effect.

Also we analytically show the possibility of the triplet spin valve effect in  F$_1$/S/F$_2$ type structures when one of the ferromagnets is weak and the other is strong. Again, the form of the dependence $T_c(\theta)$ is very sensitive to the position of the occupied spin-band in the half-metal relatively to the band in the superconductor. The triplet switching is realized if the bands in the HM and S layers coincide.

Note that the obtained results should be relevant to the artificial layered magnetic superconductors such as RuSr$_2$GdCu$_2$O$_8$ or La$_{0.7}$Ca$_{0.3}$MnO$_3$ which consist of the alternating superconducting and ferromagnetic atomic layers and also to a wide class of multilayered thin-film structures fabricated by the molecular beam epitaxy.

\acknowledgments

The authors thank A. Buzdin for suggesting the problem and continuous support, and A. Mel'nikov for the fruitful discussions. This work was supported by the Russian Science Foundation (Grant No. 15-12-10020, spin-valve effect in F/S/HM systems), and the Russian Foundation for Basic Research (Grant No. 15-02-04116, spin-valve effect in S/F/F systems).

\appendix

\section{Calculation of the anomalous Green function and the critical temperature in S/F1/F2 spin valve of atomic thickness}\label{App:AGF_SF1F2}

Expanding (\ref{AGF1}) over $t_1$ and $t_2$ up to the sixth order, we obtain the following expression for the anomalous Green function:

\begin{widetext}
\begin{multline}
\frac{\hat F^{+}}{\Delta^*}=\frac{1}{\omega_+ \omega_-}\Biggl\{\hat I +t_1^2\biggl(\frac{1}{\omega_+}\hat M_+\hat I+\frac{1}{\omega_-}\hat I\hat M_-\biggr) +t_1^2t_2^2\biggl(\frac{1}{\omega_+}\hat M_+\hat N_+ \hat M_+\hat I+\frac{1}{\omega_-}\hat I\hat M_-\hat N_- \hat M_-\biggr)+t_1^4\biggl(\frac{1}{\omega_+^2}\hat M_+^2\hat I+\frac{1}{\omega_-^2}\hat I\hat M_-^2\biggr)+\\ +t_1^2t_2^4\biggl(\frac{1}{\omega_+}(\hat M_+\hat N_+)^2  \hat M_+\hat I+\frac{1}{\omega_-}\hat I(\hat M_-\hat N_-)^2  \hat M_-\biggr)+t_1^4t_2^2\biggl(\frac{1}{\omega_+^2}\hat M_+\left(\hat M_+\hat N_+ + \hat N_+\hat M_+\right) \hat M_+\hat I+\\+\frac{1}{\omega_-^2}\hat I\hat M_-\left(\hat M_-\hat N_- + \hat N_-\hat M_-\right) \hat M_-\biggr)+t_1^6\biggl(\frac{1}{\omega_+^3}\hat M_+^3\hat I+\frac{1}{\omega_-^2}\hat I\hat M_-^3 \hat M_-+\biggr) \Biggr\},
\end{multline}
where
$\hat M_{\pm}=\left(i\omega \hat 1 \pm \hat A \right)^{-1}$, $\hat N_{\pm}=\left(i\omega \hat 1 \pm \hat B \right)^{-1}$.

Then we substitute this expression into (\ref{SCE}) and find
\begin{multline}
T_c(\theta)=T_c(0)-T_{c0}^2 \sum \limits_{\omega_n=-\infty}^{+\infty} \int  \limits_{\xi=-\infty}^{+\infty} d\xi \biggl [ -\frac{4t_1^2 t_2^2 h_1 h_2(1-\cos \theta)}{\omega_+ \omega_- (\omega_+^2-h_1^2)^2\left((\omega_+-\xi_0)^2-h_2^2\right)} \biggl(1+\frac{t_2^2(\omega_+-\xi_0)(5\omega_+^2+2h_1^2)}{2\omega_+(\omega_+^2-h_1^2)\left((\omega_+-\xi_0)^2-h_2^2\right)}+ \\+\frac{t_1^2(3\omega_+^2+h_1^2)}{\omega_+^2(\omega_+^2-h_1^2)}\biggr) -\frac{8t_1^2t_2^4 h_1^2 h_2^2 \sin^2\theta}{\omega_+ \omega_- (\omega_+^2-h_1^2)^3\left((\omega_+-\xi_0)^2-h_2^2\right)^2}\biggr]
\end{multline}
\end{widetext}

After integration we obtain (\ref{a_SFF_gf}) and (\ref{b_SFF_gf}).

\section{Gor'kov equations for the F/S/HM structure} \label{App:FSHM}

Using the same procedure as before, we obtain the following system of Gor'kov equations
\begin{gather*}
(i\omega-\xi)G+\Delta I F^{+}-t_1E^{\psi}-t_2 E^{\eta}=\hat 1, \\
(i\omega+\xi)F^+ -\Delta^* I G+t_1F^{\psi+}+t_2F^{\eta+}=0, \\
(i\omega-\hat C)E^{\psi}-t_1 G=0, \\
(i\omega+\hat C)F^{\psi+}+t_1 F^+ =0, \\
(i\omega-\hat P)E^{\eta}-t_2 G=0, \\
(i\omega+\hat P)F^{\eta+}+t_2 F^{+}=0.
\end{gather*}

The anomalous Green function can be found from this system and has the form (\ref{AGF2}) in the linear order over the gap potential $\Delta$.

We also give the exact expressions for the functions used in Sec. \ref{Sec:SFHM} which are the following
\begin{equation*}
P(x)=\frac{5+114 x^2+1053x^4+4936x^6+11472x^8+13312x^{10}}{(1+4x^2)^6(1+x^2)}.
\end{equation*}

\begin{multline*}
K(x_1,x_2)=\\\frac{P_1(x_1)+1408+P_1(x_2)+x_1^2 x_2^2 \biggl(P_2(x_1,x_2)+24-6 x_1^2-6x_2^2\biggr)}{(x_1^2+4)^2 (x_2^2+4)^2 P_2(x_1,x_2)} ,
\end{multline*}
where
$$P_1(x)=6 x^6+88 x^4+480 x^2,$$ $$P_2(x,y)=x^4-2x^2 (y^2-4)+(y^2+4)^2.$$

\end{document}